\documentclass[aps,pre,reprint,twocolumn,superscriptaddress,longbibliography,floatfix]{revtex4-2}

\usepackage{amsmath,amssymb}
\usepackage{graphicx}
\usepackage{siunitx}
\usepackage[colorlinks,linkcolor=blue,citecolor=blue,urlcolor=blue]{hyperref}

\begin{document}

\title{Fractal dimensions of paths on three-dimensional random-walk traces}

\author{Qiyuan Shi}
\affiliation{Department of Modern Physics, University of Science and Technology of China, Hefei, Anhui 230026, China}
\affiliation{Hefei National Laboratory, University of Science and Technology of China, Hefei, Anhui 230088, China}

\author{Xinyi Li}
\email{xinyili@bicmr.pku.edu.cn}
\affiliation{Beijing International Center for Mathematical Research, Peking University, Beijing 100871, China}

\author{Youjin Deng}
\email{yjdeng@ustc.edu.cn}
\affiliation{Department of Modern Physics, University of Science and Technology of China, Hefei, Anhui 230026, China}
\affiliation{Hefei National Laboratory, University of Science and Technology of China, Hefei, Anhui 230088, China}
\affiliation{Hefei National Research Center for Physical Sciences at the Microscale, University of Science and Technology of China, Hefei, Anhui 230026, China}
\date{\today}

\begin{abstract}
We study four paths that join the starting site to the final position of a simple random walk of \(N=L^2\) steps on the infinite cubic lattice. The paths are selected on the walk trace---the graph of visited sites and traversed bonds---by different rules: the undirected shortest path \(S_1\), the directed shortest path \(S_2\), the time-respecting shortest path \(S_3\), and the depth \(D_e\) of the final site in a depth-first-search tree. Monte Carlo simulations for \(L\) up to 32768 give mean lengths scaling as \(L^{d_O}\), with \(d_{S_1}=1.48005(9)\), \(d_{S_2}=1.49868(5)\), \(d_{S_3}=1.51205(5)\), and \(d_{D_e}=1.7099(7)\), so that \(d_{S_1}<d_{S_2}<d_{S_3}<d_{\rm LE}<d_{D_e}\),
where $d_{\rm LE}=1.62400(5)$ is the known loop-erased-random-walk exponent.
Imposing traversal directions increases the shortest-path exponent, and imposing time order increases it further; these constraints therefore change the scaling exponent, not only the amplitude. For every path, the standard deviation scales with the same exponent as the mean, so it is scale-invariant, unlike the visited-site mass; the rescaled length distributions collapse onto a finite-width scaling function. No analytical theory currently predicts these exponents; our results provide precise numerical benchmarks for future studies of path geometry on random-walk traces.
\end{abstract}

\maketitle

\section{Introduction}
\label{sec:intro}

Random walks were introduced by Pearson in 1905 \cite{Pearson1905}, and in
the same year Einstein related the irregular motion of suspended particles
to molecular diffusion \cite{Einstein1905}.  On a regular lattice, the
walker moves at each step to one of the neighboring sites chosen at random,
with independent steps \cite{MontrollWeiss1965,LawlerLimic2010}.  This
simple rule describes transport in amorphous solids
\cite{ScherMontroll1975}, diffusion in disordered media
\cite{HavlinBenAvraham1987}, diffusion-controlled reactions
\cite{Redner2001}, and stochastic search processes
\cite{BenichouEtAl2011}.  It is also a basic model of statistical physics,
in which a local rule produces scaling behavior that depends on the spatial
dimension \cite{Chandrasekhar1943,BouchaudGeorges1990,MetzlerKlafter2000}.

For an unbiased nearest-neighbor walk, the typical displacement after $N$
steps grows as $N^{1/2}$ \cite{Chandrasekhar1943,MontrollWeiss1965}.  The
displacement, however, does not fully characterize the trajectory.  Walks
with comparable displacements can visit different numbers of sites and
traverse different sets of bonds.  The visited sites and the traversed
bonds together form a connected graph, the trace of the walk
\cite{MovshovitzHavlin1988}.  A first visit to a site enlarges this graph,
while a later visit to a visited site closes a loop.  The structure of the
trace is therefore controlled by two properties, how often the walk returns
to the sites that it has visited and how often separated parts of the
trajectory meet, and both depend strongly on the spatial dimension $d$
\cite{ErdosTaylor1960Structure,DekeyserMaritanStella1987}.

The first property determines the mass $M$, defined as the number of distinct sites visited by the walk. The probability that the walk is back at its starting site decays as $N^{-d/2}$ after $N$ steps. In one and two dimensions this decay is slow enough that the walk returns to the starting site infinitely often, while in three and higher dimensions it returns only a finite number of times \cite{Polya1921,LawlerLimic2010}.  The mean mass grows as
\cite{DvoretzkyErdos1951}
\begin{equation}
M\propto
\begin{cases}
N/\ln N, & d=2,\\
N, & d\geq3.
\end{cases}
\label{eq:intro-mass}
\end{equation}
In two dimensions the walk returns to visited sites so often that the mass
acquires a logarithmic factor.  In three dimensions (3D) the walk is transient,
and a finite fraction $p$ of its steps reach new sites, where $p$ is the
probability that the walk never returns to its starting site.  The
fluctuations of the mass are also known.  For the simple random walk on the
cubic lattice \cite{JainPruitt1971,BassKumagai2002},
\begin{equation}
M\simeq pN,
\qquad
\operatorname{Var}(M)\simeq \frac{27p^4}{2\pi^2}N\ln N .
\label{eq:mass-3d}
\end{equation}
Here, the escape probability $p$ is related to the lattice Green function at the origin by \(p=1/G(0)\), 
where \(G(0)\) is the Watson integral for the simple cubic lattice. 
In 3D, this integral is known exactly in closed form in terms of elliptic integrals (equivalently, gamma functions), 
giving \(p\approx0.65946\ldots\)~\cite{GlasserZucker1977}. 
In higher dimensions \(d\ge4\), no comparably simple closed form is known. However, the large-\(d\) asymptotic behaviour is known analytically: \(p = 1 - 1/(2d) + \mathcal O(d^{-2})\) as \(d\to\infty\), obtained from the asymptotic expansion of the \(d\)-dimensional lattice Green function~\cite{Montroll1956}.

The second property is the intersection of separated parts of the
trajectory.  For two independent walks of $N$ steps that start at the same
site, the expected number of pairs of times at which they occupy the same
site grows as $N^{2-d/2}$ for $d<4$ and as $\ln N$ for $d=4$, and it
remains finite for $d>4$.  Hence $d_c=4$ is the upper critical dimension for
intersections of random-walk paths
\cite{DvoretzkyErdosKakutani1950,ErdosTaylor1960Intersections,
Lawler1982FourDimensions}.
It is worth noting that the same upper critical dimension also governs the \(O(n)\) spin model. This is not a coincidence: random-walk intersections correspond to the loop expansion of the \(O(n)\) model, which reduces to the self-avoiding walk as \(n\to0\), and the self-avoiding walk likewise enters its mean-field regime above four dimensions~\cite{Duplantier1988}. The intersections determine how far apart two sites are along the trace.
For $d\geq5$, a finite fraction of the steps are
cut times, after which the walk does not return to the sites visited before
\cite{ErdosTaylor1960Intersections,Lawler1996CutTimes}.  The bond traversed
at a cut time is the only link between the earlier and the later parts of
the trace, so that every path joining the two ends passes through all the
cut times.  The distance between the ends along the trace is therefore
proportional to $N$ \cite{Croydon2009Range}.  At $d=4$ the cut times are
rarer, and the distance grows as $N/\sqrt{\ln N}$
\cite{ShiraishiWatanabe2026,CroydonShiraishi2023}.  For $d<4$, the
intersections are abundant and loops exist on all length scales, so that the
distance along the trace is expected to grow as a power of $N$ smaller than
one.  Thus 3D is the only dimension below $d_c$ in which the walk is transient.

Several results are known for paths on the trace of a 3D walk.  The shortest
path between two sites of the trace has a length called the chemical
distance $\ell$, which is the minimum number of bonds in a path joining the
sites.  The sites at chemical distance $\ell$ from a given site lie at a
Euclidean distance $R$ from it that grows as $R\sim\ell^{0.68\pm0.02}$
\cite{MovshovitzHavlin1988}.  For three-dimensional Brownian motion, the
scaling limit of the walk, the corresponding shortest-path dimension is
rigorously known to exist and to lie in $(1,5/3]$ \cite{Shiraishi2026Graph}.
Besides the shortest path, chronological loop
erasure, which removes the loops of the walk in the order in which they are
formed, leaves a self-avoiding path, the loop-erased random walk (LERW).  Its
length $\ell$ grows with the Euclidean distance $R$ between its two ends as
$\ell\sim R^{d_{\rm LE}}$, where the fractal dimension $d_{\rm LE}$ exists in
3D \cite{Shiraishi2018LERW} and is numerically $d_{\rm LE}=1.62400(5)$
\cite{Wilson2010LERW}.

A path can also be selected using the information that the walk records for
each bond, the direction and the time of each traversal.  Keeping the
directions gives a directed graph, and keeping the times
as well restricts a path to traversals in increasing time order.  Such
time-respecting paths are the basic notion of distance in temporal
networks, where they can be much longer than paths on the aggregated static
graph \cite{PanSaramaki2011,HolmeSaramaki2012}.  On the trace of a random
walk, the undirected, directed, and time-respecting paths between two sites
form nested sets, and their minimum lengths are ordered in every
realization.  It is not known whether the direction and the order of the
traversals only increase the length by a constant factor or change its
scaling with $N$.

In this paper, we use Monte Carlo simulations to study paths on the trace
of simple random walks on the infinite cubic lattice, with $N=L^2$ steps up
to $L=32768$.  For each walk we measure the undirected, directed, and
time-respecting shortest paths $S_1$, $S_2$, and $S_3$ from the starting
site to the final position, and the depth $D_{\rm e}$ of the final position in a tree built by depth-first search (DFS).  For the mass $M$, we verify the exponent and the leading
amplitude in Eq.~\eqref{eq:mass-3d}, together with the logarithmic factor in
its variance.  The average length of each path follows a power law $L^{d_O}$, where $O$ stands for $S_1$, $S_2$, $S_3$, or $D_{\rm e}$, with the fractal dimensions $d_{S_1}=1.48005(9)$, $d_{S_2}=1.49868(5)$, $d_{S_3}=1.51205(5)$, and $d_{D_{\rm e}}=1.7099(7)$.  Requiring the paths to
follow the traversal directions increases the shortest-path exponent, and
imposing the time order increases it further.  The DFS exponent is larger
than those of the shortest paths, while the LERW exponent lies between those
of the time-respecting shortest path and the DFS path.  The standard
deviation of each path length grows with the same exponent as its mean, so
that the relative fluctuations remain finite and nonzero, in contrast to the
self-averaging mass.  The length distribution of each path collapses onto a
scaling function of finite width when rescaled with the fractal dimension.
We also study the cumulative distribution of the normalized path length.
Its left tail is consistent with a power law with exponent $3/d_O$, where
$d_O$ is the fractal dimension of the path.  Its right tail is consistent
with a stretched exponential with exponent $2/(2-d_O)$.

The remainder of this paper is organized as follows.
Section~\ref{sec:model} describes the model, the paths, and the algorithms.
Section~\ref{sec:results} presents the scaling of the mass and the path
lengths, their fluctuations, and their probability distributions.  The
results are discussed in Sec.~\ref{sec:discussion}.

\section{Model, observables, and algorithms}
\label{sec:model}

\begin{figure*}[t]
 \centering
 \includegraphics[width=\textwidth]{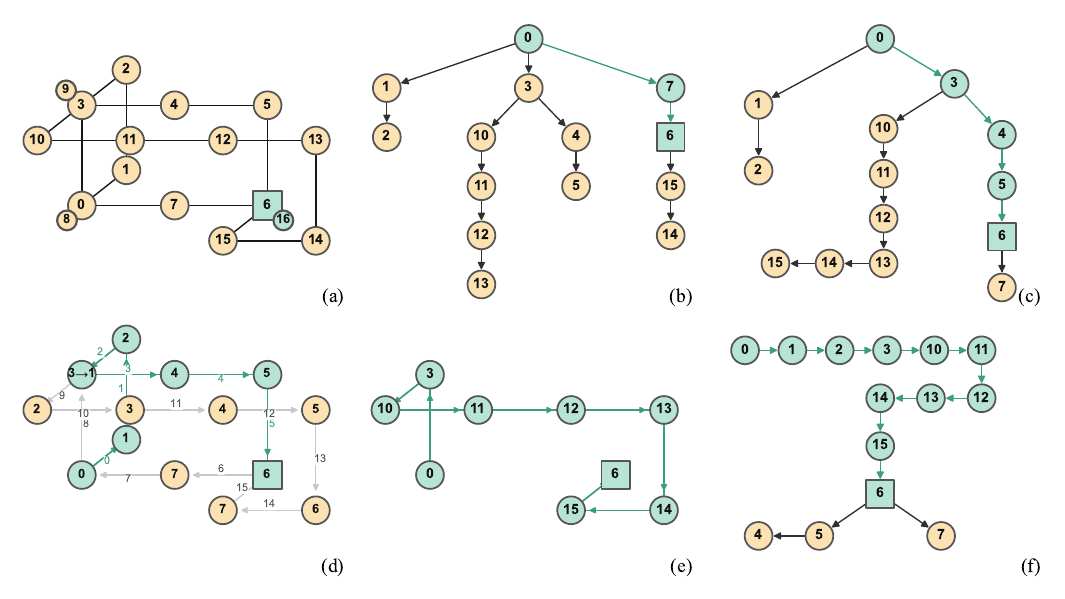}
 \caption{Schematic illustration of paths constructed from a 16-step random
 walk on the cubic lattice.  (a) One realization of the walk.  The number
 $t$ identifies the site $X_t$ occupied at step $t$, for $t=0,\ldots,16$.
 Each distinct site is labeled by its first-visit time, and an attached
 smaller circle records a later visit to the same site.  The walk starts at
 the circle labeled 0.  Its final position at $t=16$ is the site first
 visited at $t=6$, which is marked by a square and joined to the attached
 circle labeled 16.  The walk visits $\mathcal M=14$ distinct sites.  Panels
 (b)--(f) are constructed from the walk in panel (a).  In panels (b), (c),
 (e), and (f), each site is identified by its first-visit time.  The square
 marks the final site $X_{16}$ in each panel, and in panels (b)--(f) the
 selected path is highlighted in green.  (b) BFS on the undirected bond
 graph $G_1$ formed by the walk produces a search tree whose branch from the
 root to the final site is $0\to7\to6$, so $\mathcal S_1=2$.  (c) BFS on the
 directed graph $G_2$ formed from the traversal directions of the same walk
 gives the search tree with branch $0\to3\to4\to5\to6$, so
 $\mathcal S_2=4$.  (d) Each directed traversal $X_t\to X_{t+1}$ is labeled
 by its time $t$.  The numbers inside each node record successive updates of
 $d_t(v)$, the time-respecting distance from $X_0$, according to
 Eq.~\eqref{eq:temporal-update}.  The path in panel (c) uses the traversal at
 time 8 before those at times 3, 4, and 5 and hence violates chronological
 order.  The green path in panel (d) instead uses times 0 through 5 and
 gives $\mathcal S_3=6$.  (e) For comparison, chronological loop erasure
 leaves a path of length 8.  (f) DFS of $G_1$ for the same walk gives the
 search tree whose highlighted branch from the root to the final site has
 length $\mathcal D_{\rm e}=10$.}
 \label{fig:path-definitions}
\end{figure*}

We simulate a discrete-time simple random walk $X_t$ on the infinite
cubic lattice with unit lattice spacing.  The walk starts at the origin,
$X_0=\boldsymbol 0$, and at each step it moves with equal probability to
one of the six nearest neighbors.  We simulate walks of different lengths
and take measurements at the times $N=L^2$ along each walk, with
$16\leq L\leq32768$. The sample size depends on $L$, decreasing from
order $10^7$ at the smallest sizes to order $10^5$ at the largest.
Since the typical displacement after $N$ steps is of
order $N^{1/2}$, $L$ is the diffusive length scale of the walk.  The lattice
is infinite, and no boundary is imposed.  We measure the following
observables, which are illustrated in Fig.~\ref{fig:path-definitions} for a
walk of 16 steps.

\emph{Mass $\mathcal M$.}
The mass is the number of distinct sites visited up to time $N$,
$\mathcal M=|V_N|$ with $V_N=\{X_t:0\leq t\leq N\}$.  In
Fig.~\ref{fig:path-definitions}(a), the 17 positions include three revisits,
and $\mathcal M=14$.

\emph{Undirected shortest path $\mathcal S_1$.}
The visited sites and the traversed bonds
$E_N=\{\{X_t,X_{t+1}\}:0\leq t<N\}$ form the connected graph
$G_1=(V_N,E_N)$.  The length of a path is the number of bonds that it
contains, and the distance between two sites is the minimum length over all
paths joining them.  The length $\mathcal S_1$ is the distance between $X_0$
and $X_N$ on $G_1$.  It is obtained by a breadth-first search (BFS) from
$X_0$, which explores $G_1$ layer by layer, so that the layer in which $X_N$
first appears gives $\mathcal S_1$ [Fig.~\ref{fig:path-definitions}(b)].

\emph{Directed shortest path $\mathcal S_2$.}
Assigning the observed traversal directions to the bonds of $G_1$ gives the
directed graph $G_2$, in which a bond traversed in both directions carries
two edges.  The length $\mathcal S_2$ is the minimum number of directed
edges in a path from $X_0$ to $X_N$ on $G_2$.  It is computed by BFS along the edge
directions [Fig.~\ref{fig:path-definitions}(c)].

\emph{Time-respecting shortest path $\mathcal S_3$.}
A time-respecting path from $X_0$ to $X_N$ is a sequence of traversals
whose time labels $t$ strictly increase.  Consecutive traversals in the
path need not be consecutive steps of the walk.  The length
$\mathcal S_3$ is the minimum number of traversals in such a path.  It can
be computed during the walk.  Let $d_t(v)$ be the minimum length of a
time-respecting path from $X_0$ to $v$ that uses only the first $t$
traversals, with $d_0(X_0)=0$ and $d_0(v)=\infty$ otherwise.  The
traversal $X_t\to X_{t+1}$ can extend a shortest path to $X_t$, so that
\begin{equation}
 d_{t+1}(X_{t+1})=
 \min\!\bigl\{d_t(X_{t+1}),\,d_t(X_t)+1\bigr\},
 \label{eq:temporal-update}
\end{equation}
while all other distances are unchanged.  The measured length is
$\mathcal S_3=d_N(X_N)$ [Fig.~\ref{fig:path-definitions}(d)].  Since every
time-respecting path is a path on $G_2$, and every path on $G_2$ is a path
on $G_1$, the three lengths satisfy
$\mathcal S_1\leq\mathcal S_2\leq\mathcal S_3$ in every realization.  The
loop-erased walk of Fig.~\ref{fig:path-definitions}(e) consists of
traversals in increasing time order, and its length is therefore never
smaller than $\mathcal S_3$.

\emph{DFS depth $\mathcal D_{\rm e}$.}
The BFS gives the shortest paths on $G_1$. The other basic graph traversal, depth-first search (DFS), also gives a path from $X_0$ to $X_N$. Instead of exploring layer by layer, it follows each branch as far as possible before backtracking, and the DFS paths are fractal in critical clusters of the $O(n)$ loop model and of percolation \cite{ShiDengLi2025DFS}.  Starting at $X_0$, the
DFS moves to an unvisited neighbor, chosen at random among the available
ones, whenever one exists, and backtracks otherwise.  The bonds used to reach
unvisited sites form a spanning tree rooted at $X_0$.  The length
$\mathcal D_{\rm e}$ is the number of bonds on the tree path from the root
to $X_N$ [Fig.~\ref{fig:path-definitions}(f)].

In the following, $\mathcal O$ denotes any of the observables $\mathcal M$, $\mathcal S_1$, $\mathcal S_2$, $\mathcal S_3$, and $\mathcal D_{\rm e}$, and $O=\langle\mathcal O\rangle$ denotes its mean over independent walks, namely $M$, $S_1$, $S_2$, $S_3$, and $D_{\rm e}$. We also measure the fluctuation
\begin{equation}
 \chi_O(L)=\frac{\langle\mathcal O^2\rangle-
 \langle\mathcal O\rangle^2}{L^2},
 \label{eq:fluctuation}
\end{equation}
and the histogram of each observable at every measurement time.

\section{Numerical results}
\label{sec:results}

We first test the simulations against the exact results for the mass, and
then determine the fractal dimensions of the paths and the scaling of their
fluctuations and distributions.  To determine the scaling exponents, we fit the Monte Carlo data of a quantity $O$ to the finite-size scaling ansatz
\begin{equation}
 O(L)=L^{d_O}\left(a+bL^{-\omega}\right).
 \label{eq:power-fit}
\end{equation}
Here $d_O$ is the scaling exponent, and $bL^{-\omega}$ describes the
leading finite-size correction.  The correction term is omitted when it is
not required by the data.  We fit the data for $L\geq L_{\min}$ and assess
the goodness of fit using $\chi^2/\mathrm{DF}$.  Here $\chi^2$ measures the
fit residuals, and DF is the number of degrees of freedom.  Fits with
$\chi^2/\mathrm{DF}\simeq1$ are considered acceptable.  The final estimates
are obtained from fits that remain stable when $L_{\min}$ and the correction
exponent $\omega$ are varied.  Table~\ref{tab:exponent-summary} summarizes
the resulting exponents.

\begin{table}[t]
\caption{Fractal dimensions $d_O$ of the mean lengths and exponents
$d_{\chi_O}$ of the fluctuations for the 3D random-walk trace.  The mass exponent $d_M=2$ follows from Eq.~\eqref{eq:mass-3d}.
The last column gives $2d_O-2$ for comparison with $d_{\chi_O}$.}
\label{tab:exponent-summary}
\centering
\begin{ruledtabular}
\begin{tabular}{c l l l}
$O$ & $d_O$ & $d_{\chi_O}$ & $2d_O-2$\\
\hline
$M$ & $2$ & & \\
$S_1$ & $1.48005(9)$ & $0.9601(4)$ & $0.9601(2)$\\
$S_2$ & $1.49868(5)$ & $0.9977(2)$ & $0.9974(1)$\\
$S_3$ & $1.51205(5)$ & $1.0241(2)$ & $1.0241(1)$\\
$D_{\rm e}$ & $1.7099(7)$ & $1.4194(3)$ & $1.4198(14)$\\
\end{tabular}
\end{ruledtabular}
\end{table}

\subsection{Mass of the walk}
\label{sec:range}

\begin{figure}[t]
 \centering
 \includegraphics[width=\columnwidth]{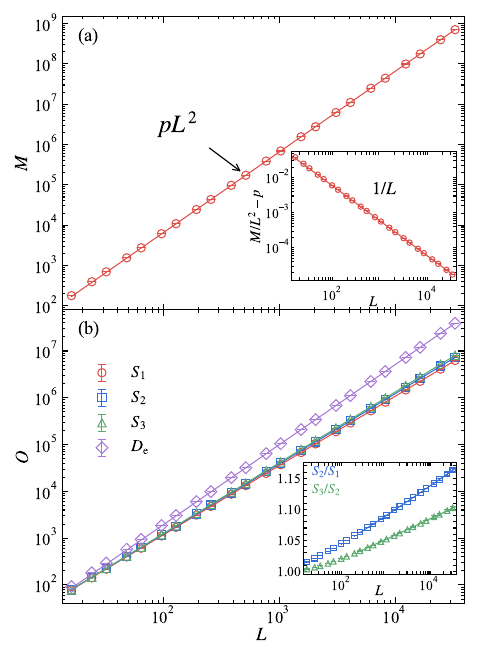}
 \caption{Mean mass and mean path lengths of 3D random walks against the diffusive length $L$.  Open symbols are Monte Carlo
 data.  Error bars are smaller than the symbol sizes in both main panels.
 (a) The mass $M$ and the exact asymptote $pL^2$ (solid line), where
 $p$ is the escape probability.  The inset shows $M/L^2-p$, and the solid
 line is proportional to $L^{-1}$.
 (b) The mean path lengths $S_1$, $S_2$, $S_3$, and $D_{\rm e}$.  The solid
 lines have slopes given by the exponents $d_O$ in
 Table~\ref{tab:exponent-summary}.  The inset shows the
 ratios $S_2/S_1$ and $S_3/S_2$ on a logarithmic $L$ axis, and the solid
 lines are proportional to $L^{d_{S_2}-d_{S_1}}$ and
 $L^{d_{S_3}-d_{S_2}}$.}
 \label{fig:mean-scaling}
\end{figure}

The mass has exact asymptotics, so it provides a test of the simulations
before we turn to the paths.  At $N=L^2$, Eq.~\eqref{eq:mass-3d} gives
$M\simeq pL^2$, where $p\approx0.65946\ldots$ is the escape probability of
the cubic lattice \cite{GlasserZucker1977}.  We fit $M$ to
Eq.~\eqref{eq:power-fit}.  The fits give $d_M=1.999999(1)$, in agreement with the exact exponent $2$, together with $a=0.65947(1)$, which agrees with $p$, and the correction exponent $\omega=1.005(2)$.  The data in
Fig.~\ref{fig:mean-scaling}(a) follow this law, and the inset shows that
$M/L^2-p$ decays as $L^{-1}$.

\begin{figure}[t]
 \centering
 \includegraphics[width=\columnwidth]{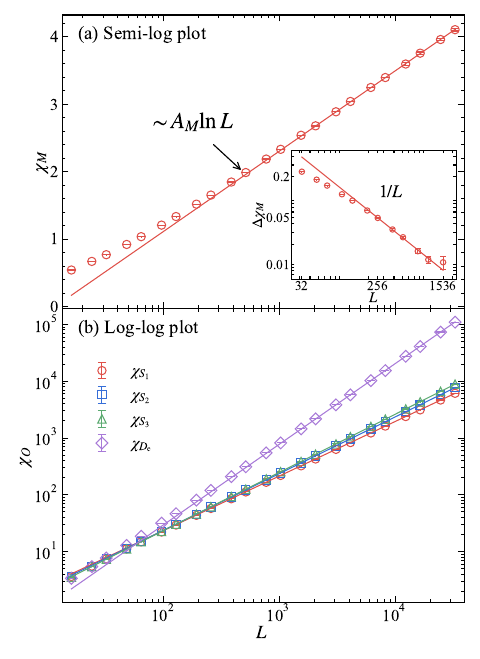}
 \caption{Fluctuations $\chi_O$, Eq.~\eqref{eq:fluctuation}, of the mass
 and the path lengths.  Open symbols are Monte Carlo data.  Error bars are
 smaller than the symbol sizes in both main panels.  (a) The mass
 fluctuation $\chi_M$ on a logarithmic horizontal axis.  The solid line is
 $A_M\ln L+B$, with the exact coefficient $A_M$ of
 Eq.~\eqref{eq:mass-logarithm} and the constant $B$ from a fit to
 Eq.~\eqref{eq:mass-fluctuation-fit}.  The inset shows the remaining
 correction $\Delta\chi_M=\chi_M-A\ln L-B$.  The solid line indicates its
 $L^{-1}$ decay.  (b) The
 path-length fluctuations $\chi_{S_1}$, $\chi_{S_2}$, $\chi_{S_3}$, and
 $\chi_{D_{\rm e}}$.  The solid lines are power laws with the exponents
 $d_{\chi_O}$ in Table~\ref{tab:exponent-summary}.}
 \label{fig:fluctuation-scaling}
\end{figure}

The fluctuation of the mass is tested in the same way.  Let $\sigma_M$ be
the standard deviation of $\mathcal M$, so that $\chi_M=\sigma_M^2/L^2$.
Equations~\eqref{eq:mass-3d} and \eqref{eq:fluctuation} give
\begin{equation}
 \chi_M(L)\simeq A_M\ln L,
 \qquad A_M=\frac{27p^4}{\pi^2}\approx0.51740\ldots .
 \label{eq:mass-logarithm}
\end{equation}
We fit $\chi_M$ to
\begin{equation}
 \chi_M(L)=A\ln L+B+CL^{-\omega}.
 \label{eq:mass-fluctuation-fit}
\end{equation}
Here $B$ is a constant term, and $CL^{-\omega}$ describes the leading
power-law correction.  The fits give $\omega$ close to $1$.  With
$\omega=1$ fixed, the coefficient is stable at $A=0.518(3)$, in agreement
with $A_M$.  The logarithmic growth is shown in
Fig.~\ref{fig:fluctuation-scaling}(a), and the inset shows that the
remaining correction decays as $L^{-1}$.  The standard deviation and the
relative fluctuation of the mass therefore behave as
\begin{equation}
 \sigma_M\propto L\sqrt{\ln L},
 \qquad
 \frac{\sigma_M}{M}\propto\frac{\sqrt{\ln L}}{L}\longrightarrow0 .
 \label{eq:mass-relative}
\end{equation}
The mass is self-averaging, and its relative fluctuation vanishes as
$L\to\infty$.  This serves as the reference for the path lengths, whose
fluctuations are studied below.

\subsection{Fractal dimensions of the paths}
\label{sec:mean-scaling}

\begin{table}[t]
\caption{Fits of the mean path lengths $S_1$, $S_2$, $S_3$, and
$D_{\rm e}$ to Eq.~\eqref{eq:power-fit}.  Parameters shown without error
bars are fixed in the fits.}
\label{tab:mean-fits}
\centering
\setlength{\tabcolsep}{1pt}
\begin{ruledtabular}
\begin{tabular}{llllS[table-format=-1.4(1),group-digits=false]ll}
$O$ & $L_{\min}$ & $d_O$ & $a$ & \multicolumn{1}{c}{$b$} & $\omega$ & $\chi^2/\mathrm{DF}$\\
\hline
$S_1$ & 16  & 1.48012(5) & 1.3008(4) & -0.59(2) & 1.10(1) & 13.9/19\\
      & 32  & 1.48006(7) & 1.3015(7) & -0.51(6) & 1.05(4) & 12.9/17\\
      & 64  & 1.4801(1) & 1.301(1) & -0.6(3) & 1.1(1) & 12.6/15\\
      & 32  & 1.47999(3) & 1.3023(3) & -0.440(6) & 1 & 13.9/18\\
      & 64  & 1.48001(5) & 1.3020(4) & -0.43(1) & 1 & 13.3/16\\
      & 128 & 1.48006(7) & 1.3016(7) & -0.40(4) & 1 & 12.7/14\\
\hline
$S_2$ & 16 & 1.49867(4) & 1.2472(3) & -0.60(3) & 1.23(2) & 13.7/19\\
      & 32 & 1.49867(6) & 1.2473(5) & -0.6(1) & 1.22(6) & 13.4/17\\
      & 64 & 1.49870(8) & 1.2469(8) & -1.0(8) & 1.4(2) & 13.0/15\\
      & 16 & 1.49869(2) & 1.2470(1) & -0.625(4) & $5/4$ & 14.1/20\\
      & 32 & 1.49869(3) & 1.2471(2) & -0.63(1) & $5/4$ & 13.6/18\\
      & 64 & 1.49867(4) & 1.2473(3) & -0.65(3) & $5/4$ & 13.2/16\\
\hline
$S_3$ & 16  & 1.51200(3) & 1.1962(2) & -1.4(2) & 1.85(7) & 19.8/19\\
      & 24  & 1.51204(3) & 1.1959(2) & -4(2) & 2.2(1) & 16.0/18\\
      & 32  & 1.51205(3) & 1.1958(3) & -7(8) & 2.3(3) & 15.7/17\\
      & 32  & 1.51202(2) & 1.1960(2) & -2.3(1) & 2 & 16.6/18\\
      & 64  & 1.51204(4) & 1.1959(3) & -1.8(6) & 2 & 16.1/16\\
      & 128 & 1.51206(3) & 1.1957(2) & & & 14.4/15\\
      & 256 & 1.51205(4) & 1.1958(3) & & & 11.4/13\\
\hline
$D_{\rm e}$ & 1024 & 1.7100(3) & 0.742(2) & 45(104) & 1.4(4) & 2.2/7\\
            & 1024 & 1.7099(1) & 0.7422(7) & 77(5) & $3/2$ & 2.3/8\\
            & 2048 & 1.7099(2) & 0.742(1) & 78(26) & $3/2$ & 2.3/6\\
            & 4096 & 1.7100(5) & 0.741(3) & 95(126) & $3/2$ & 2.1/4\\
            & 2048 & 1.7093(1) & 0.7462(8) & & & 5.6/7\\
            & 4096 & 1.7097(2) & 0.744(1) & & & 2.4/5\\
\end{tabular}
\end{ruledtabular}
\end{table}

\begin{figure}[t]
 \centering
 \includegraphics[width=\columnwidth]{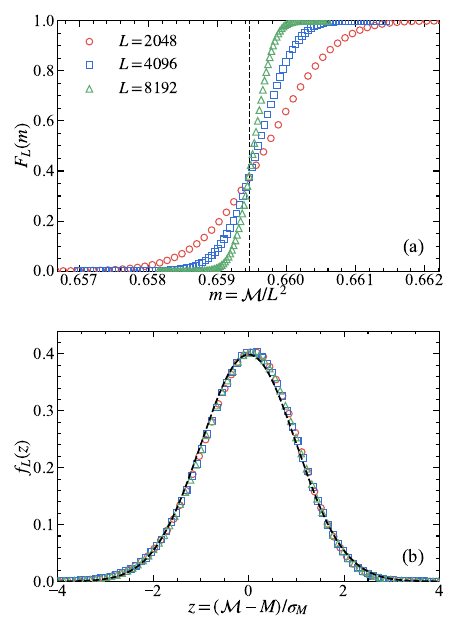}
 \caption{Distributions of the mass for $L=2048$, 4096, and 8192.
 (a) Cumulative distribution $F_L(m)$ of $m=\mathcal M/L^2$.  The vertical
 dashed line marks the escape probability $p$.  (b) Probability density
 $f_L(z)$ of the standardized mass $z=(\mathcal M-M)/\sigma_M$.  The dashed
 curve is the standard Gaussian density.}
 \label{fig:mass-distributions}
\end{figure}

The mass counts the visited sites, while the path lengths depend on how the walk connects them. We first consider the undirected shortest path $S_1$ on $G_1$.  The mean lengths in
Fig.~\ref{fig:mean-scaling}(b) grow as power laws of $L$, and we fit them to
Eq.~\eqref{eq:power-fit}.  For $S_1$, the fit with free $\omega$ in
Table~\ref{tab:mean-fits} gives $\omega$ close to $1$.  With $\omega=1$
fixed, the fits are stable when $L_{\min}$ is varied, and we estimate
$d_{S_1}=1.48005(9)$.  The same procedure is used for the other paths.  For
$S_2$ the free fits give $\omega$ near $5/4$, and fixing $\omega=5/4$ gives
$d_{S_2}=1.49868(5)$.  For $S_3$ we obtain $d_{S_3}=1.51205(5)$.  The three shortest paths have distinct exponents,
$d_{S_1}<d_{S_2}<d_{S_3}$.  Requiring a path to follow the traversal
directions, and then to follow the time order, therefore increases the
fractal dimension of the shortest path.

The exponents differ by only $0.019$ and $0.013$ and the amplitudes are close, so the three curves nearly overlap in Fig.~\ref{fig:mean-scaling}(b).  The inset shows the ratios
$S_2/S_1$ and $S_3/S_2$.  If the restrictions changed only the amplitudes,
the ratios would approach constants.  Instead, they keep increasing as
powers of $L$ with the exponent differences, which supports the distinct
exponents obtained from the fits.

The DFS branch is also a path on $G_1$, so $\mathcal D_{\rm e}\geq
\mathcal S_1$ in every realization.  It is not required to be short, and
its fractal dimension $d_{D_{\rm e}}=1.7099(7)$ is much larger than those of
the shortest paths.    The path lengths therefore grow
faster than the displacement $L$ and more slowly than the number of steps
$N=L^2$, with $1<d_O<d_M=2$ for all four paths.

These paths can be compared with the chemical distance and the LERW in 3D.
The relation $R\sim\ell^{0.68\pm0.02}$ between the Euclidean distance $R$ and
the chemical distance $\ell$ \cite{MovshovitzHavlin1988} is equivalent to
$\ell\sim R^{d_\ell}$ with $d_\ell=1/0.68=1.47(4)$.  The undirected shortest
path $S_1$ is the chemical distance between $X_0$ and $X_N$, whose
Euclidean distance is of order $L$, so that $S_1\sim L^{d_{S_1}}$ has the same
exponent.  Our result $d_{S_1}=1.48005(9)$ is consistent with $d_\ell$, with a
higher precision, and falls inside the interval $(1,5/3]$ in which the
shortest-path dimension for three-dimensional Brownian motion is rigorously
known to lie \cite{Shiraishi2026Graph}.  The LERW is a time-respecting path on the trace, so that its length is never
smaller than $\mathcal S_3$, which is consistent with
$d_{S_3}<d_{\rm LE}=1.62400(5)$ \cite{Wilson2010LERW}, measured away from the
starting point of the loop-erased path.  The shortest paths use the loops of
the trace as shortcuts, whereas the LERW is fixed by the history of the
walk.

The $O(n)$ loop model \cite{DOMANY1981279} provides a one-parameter family
of random paths, whose loops are characterized by a fractal dimension that
depends on the loop weight $n$.  At $n=-2$ the loop model corresponds to the
LERW through the field-theoretic mapping of $\phi^4$ theory
\cite{WieseFedorenko2019LERW}, and at $n=0$ it reduces to the
self-avoiding walk (SAW), with $d_{\rm SAW}=1/\nu=1.70185$
\cite{ClisbyDuenweg2016SAW}.  For $n=1$, 2, and 3, the loop dimensions are
$1.734(4)$, $1.765(3)$, and $1.795(3)$ \cite{LiuEtAl2012Loops}, which
increase with $n$ over these values.  The fractal dimensions of the three
shortest paths are smaller than all of the dimensions listed.  The DFS
dimension is slightly larger than $d_{\rm SAW}$ and smaller than the loop
dimension at $n=1$.  A DFS branch is self-avoiding, but it is selected by
the search rule on the trace and not with the uniform weight of the SAW.
Altogether,
\begin{equation}
 d_{S_1}<d_{S_2}<d_{S_3}<d_{\rm LE}<d_{\rm SAW}<d_{D_{\rm e}}<d_M=2.
 \label{eq:exponent-ordering}
\end{equation}

\begin{figure*}[t]
 \centering
 \includegraphics[width=\textwidth]{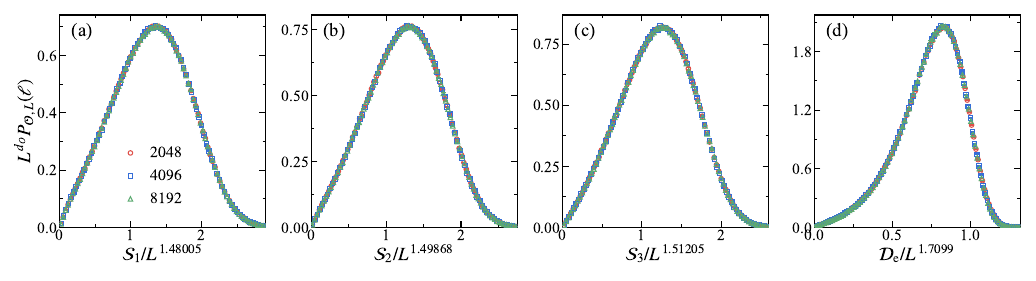}
 \caption{Scaling collapse of the path-length probability densities for
 $L=2048$, 4096, and 8192.  Panels (a)--(d) show $\mathcal S_1$,
 $\mathcal S_2$, $\mathcal S_3$, and $\mathcal D_{\rm e}$, respectively.  The
 densities are plotted as $L^{d_O}P_{\mathcal O,L}(\ell)$ against
 $\ell/L^{d_O}$, as in Eq.~\eqref{eq:path-distribution-scaling}, where
 $\ell$ is the path length and $d_O$ is taken from
 Table~\ref{tab:exponent-summary}.}
 \label{fig:path-scaling-functions}
\end{figure*}

\begin{figure*}[t]
 \centering
 \includegraphics[width=\textwidth]{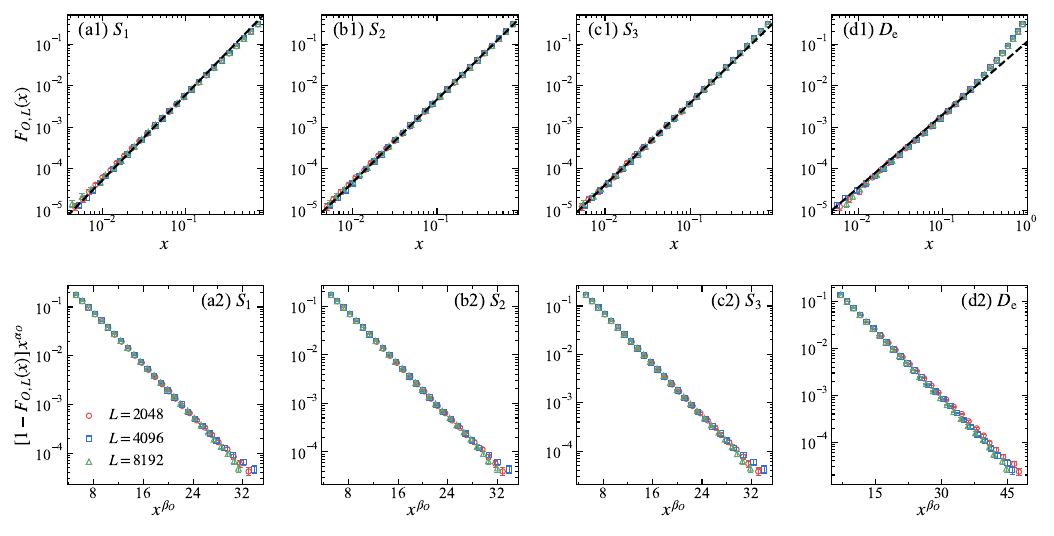}
 \caption{Tails of the path-length distributions for $L=2048$, 4096, and
 8192, with $x=\mathcal O/O$ and $F_{O,L}(x)=\Pr(\mathcal O/O\leq x)$.
 Panels (a1)--(d1) show the left tails $F_{O,L}(x)$ of $\mathcal S_1$,
 $\mathcal S_2$, $\mathcal S_3$, and $\mathcal D_{\rm e}$ on logarithmic
 axes, where the dashed lines have slope $3/d_O$, as in
 Eq.~\eqref{eq:path-left-tail}.  Panels (a2)--(d2) show the right tails on
 semilogarithmic axes,
 with $[1-F_{O,L}(x)]x^{\alpha_O}$ plotted against $x^{\beta_O}$, where
 $\alpha_O$ and $\beta_O$ are given in Eq.~\eqref{eq:path-right-tail}.}
 \label{fig:path-tails}
\end{figure*}

\subsection{Fluctuations of the path lengths}
\label{sec:fluctuations}

\begin{table}[t]
\caption{Fits of the path-length fluctuations $\chi_O$ to
Eq.~\eqref{eq:power-fit} with $O=\chi_O$ and $d_O=d_{\chi_O}$.  Parameters
shown without error bars are fixed in the fits.}
\label{tab:path-fluctuation-fits}
\centering
\setlength{\tabcolsep}{1pt}
\begin{ruledtabular}
\begin{tabular}{llllS[table-format=-1.4(1),group-digits=false]ll}
$O$ & $L_{\min}$ & $d_{\chi_O}$ & $a$ & \multicolumn{1}{c}{$b$} & $\omega$ & $\chi^2/\mathrm{DF}$\\
\hline
$S_1$ & 16  & 0.9599(2) & 0.2840(6) & -0.268(7) & 0.81(1) & 25.4/19\\
      & 32  & 0.9605(3) & 0.2825(8) & -0.32(2) & 0.87(3) & 19.0/17\\
      & 16  & 0.9598(1) & 0.2842(2) & -0.265(1) & $4/5$ & 25.7/20\\
      & 32  & 0.9598(1) & 0.2843(3) & -0.265(3) & $4/5$ & 24.3/18\\
      & 64  & 0.9602(2) & 0.2834(4) & -0.249(6) & $4/5$ & 16.3/16\\
\hline
$S_2$ & 16  & 0.9971(2) & 0.2426(5) & -0.180(8) & 0.87(2) & 25.8/19\\
      & 32  & 0.9977(2) & 0.2413(5) & -0.25(3) & 0.99(4) & 17.5/17\\
      & 32  & 0.9977(1) & 0.2412(2) & -0.260(4) & 1 & 17.6/18\\
      & 64  & 0.9978(1) & 0.2411(2) & -0.25(1) & 1 & 15.9/16\\
      & 128 & 0.9977(2) & 0.2413(4) & -0.26(2) & 1 & 15.7/14\\
\hline
$S_3$ & 128 & 1.0237(1) & 0.2116(1) & & & 22.6/15\\
      & 256 & 1.0240(1) & 0.2111(2) & & & 14.0/13\\
      & 512 & 1.0241(2) & 0.2109(3) & & & 13.2/11\\
\hline
$D_{\rm e}$ & 16 & 1.4190(3) & 0.0428(1) & 0.188(1) & 0.754(3) & 21.1/19\\
             & 32 & 1.4196(5) & 0.0426(1) & 0.183(3) & 0.745(6) & 18.3/17\\
             & 16 & 1.4195(1) & 0.04268(5) & 0.1858(2) & $3/4$ & 23.1/20\\
             & 32 & 1.4193(2) & 0.04276(6) & 0.1850(4) & $3/4$ & 18.8/18\\
             & 64 & 1.4194(2) & 0.04271(9) & 0.1856(9) & $3/4$ & 17.7/16\\
\end{tabular}
\end{ruledtabular}
\end{table}

We next study the scaling of the fluctuations $\chi_O$ of the path lengths.
They also follow power laws, as shown in
Fig.~\ref{fig:fluctuation-scaling}(b), and we fit them to Eq.~\eqref{eq:power-fit} with $O$ replaced by $\chi_O$.  The fits for $S_1$ are first
performed with free $\omega$, which gives $\omega$ close to $4/5$, and then
with $\omega=4/5$ fixed, which gives a stable estimate.  The other paths are
treated in the same way, with the fits listed in
Table~\ref{tab:path-fluctuation-fits}.  The resulting exponents are
$d_{\chi_{S_1}}=0.9601(4)$, $d_{\chi_{S_2}}=0.9977(2)$,
$d_{\chi_{S_3}}=1.0241(2)$, and $d_{\chi_{D_{\rm e}}}=1.4194(3)$.  Comparing
them with the path dimensions in Table~\ref{tab:exponent-summary}, we find
\begin{equation}
 d_{\chi_O}=2d_O-2
 \label{eq:fluctuation-relation}
\end{equation}
within the errors.  Let $\sigma_O$ be the standard deviation of the path
length $\mathcal O$, so that $\chi_O=\sigma_O^2/L^2$ by
Eq.~\eqref{eq:fluctuation}.  Equation~\eqref{eq:fluctuation-relation} then
gives
\begin{equation}
 \chi_O=\frac{\sigma_O^2}{L^2}\propto L^{2d_O-2},
 \qquad \sigma_O\propto L^{d_O}.
 \label{eq:standard-deviation}
\end{equation}
The standard deviation and the average of each path length thus grow with
the same exponent, and the relative standard deviation $\sigma_O/O$
approaches a finite, nonzero constant.  Fluctuations between walks
therefore remain of the same order as the average path length at all
sizes, and the length of a path at one size is statistically similar to that
at another after dividing by $L^{d_O}$.  These results for the mean and the
variance are consistent with scale invariance of the path lengths over the
measured sizes.  They differ from the mass, whose relative fluctuation
vanishes as in Eq.~\eqref{eq:mass-relative}, so that the path lengths are
not self-averaging.  A finite relative fluctuation does not by itself fix
the shape of the distribution, which we examine next.

\subsection{Probability distributions}
\label{sec:distributions}

We next examine the probability distributions of the mass and the path
lengths.  For the mass, $M\to pL^2$ and $\sigma_M/M\to0$ by
Eq.~\eqref{eq:mass-relative}, so that the probability density of
$m=\mathcal M/L^2$ approaches $\delta(m-p)$ as $L\to\infty$.  A
$\delta$ function is equivalent to a step in the cumulative distribution
$F_L(m)=\Pr(\mathcal M/L^2\leq m)$, which is easier to display.  As shown in
Fig.~\ref{fig:mass-distributions}(a), the curves for finite $L$ become
steeper near $m=p$ as $L$ increases and approach a step.  To examine the
fluctuations around the mean, we use the standardized variable
$z=(\mathcal M-M)/\sigma_M$ and denote its probability density by $f_L(z)$.
The central limit theorem for the mass predicts that $z$ follows the
standard Gaussian distribution as $L\to\infty$
\cite{JainPruitt1971,BassKumagai2002}.  The densities in
Fig.~\ref{fig:mass-distributions}(b) collapse well for different $L$ and
approach this Gaussian distribution as $L$ increases.

In contrast, the path lengths have a finite relative fluctuation, so that
the rescaled variable $\mathcal O/L^{d_O}$ has a distribution of finite
width.  Let $P_{\mathcal O,L}(\ell)$ denote the probability density of the path length $\mathcal O$. We examine the scaling form
\begin{equation}
 P_{\mathcal O,L}(\ell)\simeq L^{-d_O}f_O\!\left(\ell/L^{d_O}\right),
 \label{eq:path-distribution-scaling}
\end{equation}
where $f_O$ is the scaling function and $d_O$ is the fractal dimension
determined in Sec.~\ref{sec:mean-scaling}.  The finite width alone does not fix this form, and we examine whether the same scale $L^{d_O}$ describes the full distribution.  Figure~\ref{fig:path-scaling-functions} shows
$L^{d_O}P_{\mathcal O,L}(\ell)$ against $\ell/L^{d_O}$ for different $L$,
using the exponents in Table~\ref{tab:exponent-summary}.  The distributions
of all four paths are unimodal, and the data for different $L$ collapse well
onto a single curve for each path.  Thus the exponent obtained from the
average length also describes the scaling of the full distribution, which is
consistent with the common growth exponent of the average length and the
standard deviation in Sec.~\ref{sec:fluctuations}.  The scaling functions of
the three shortest paths are similar in shape, while that of the DFS branch
is narrower and more asymmetric.

\begin{table}[t]
\caption{Tail exponents of the path-length distributions at $L=8192$.
$\theta_O$ is fitted to Eq.~\eqref{eq:path-left-tail}, and $\beta_O$ is
fitted to Eq.~\eqref{eq:path-right-tail} with $\alpha_O=2/d_O$.  The
predictions $3/d_O$ and $2/(2-d_O)$ use the exponents in
Table~\ref{tab:exponent-summary}.}
\label{tab:tails}
\centering
\begin{ruledtabular}
\begin{tabular}{c c c c c}
$O$ & $\theta_O$ & $3/d_O$ & $\beta_O$ & $2/(2-d_O)$\\
\hline
$S_1$ & $1.998(30)$ & $2.027$ & $3.91(10)$ & $3.85$\\
$S_2$ & $1.992(33)$ & $2.002$ & $4.02(11)$ & $3.99$\\
$S_3$ & $1.996(34)$ & $1.984$ & $4.12(12)$ & $4.10$\\
$D_{\rm e}$ & $1.814(28)$ & $1.754$ & $6.69(17)$ & $6.89$\\
\end{tabular}
\end{ruledtabular}
\end{table}

The tails of the scaling functions \cite{ZhuEtAl2026Tails} describe the
probabilities of unusually short and unusually long paths.  We use the
normalized path length $x=\mathcal O/O$ and its cumulative
distribution $F_{O,L}(x)=\Pr(\mathcal O/O\leq x)$.  For the right tail, a
path cannot be longer than the number of steps $N$, so that $x$ is bounded by
$N/O\sim N^{1-d_O/2}$.  If the probability of a path of length of order $N$
is exponentially small in $N$, a stretched exponential $\exp(-cx^{\beta})$
matches this bound when $x^{\beta}\sim N$ at $x\sim N^{1-d_O/2}$, which gives
$\beta=2/(2-d_O)$, as in the argument of Fisher for the end-to-end distance
of polymers \cite{Fisher1966Shape}.  We therefore expect
\begin{equation}
 1-F_{O,L}(x)\sim x^{-\alpha_O}\exp\!\left(-cx^{\beta_O}\right),
 \qquad x\to\infty ,
 \label{eq:path-right-tail}
\end{equation}
with $\beta_O=2/(2-d_O)$, and we take $\alpha_O=2/d_O$ for the algebraic
prefactor following Ref.~\cite{ZhuEtAl2026Tails}.

For the left tail, a short path is likely to occur when the walk ends close to
its starting point, at a distance $r\ll L$.  If the path length scales as
$r^{d_O}$ with the distance $r$ between its two ends, then
$x\sim(r/L)^{d_O}$, while the probability that the walk ends within $r$ of
its starting point is $(r/L)^d$ with $d=3$.  This suggests an algebraic left
tail,
\begin{equation}
 F_{O,L}(x)\sim x^{\theta_O},\qquad \theta_O=\frac{3}{d_O},
 \qquad x\to0 .
 \label{eq:path-left-tail}
\end{equation}

Figure~\ref{fig:path-tails} shows the two tails.  The left tails in panels
(a1)--(d1) are straight on logarithmic axes and follow the slope $3/d_O$.  The
fitted powers $\theta_O$ in Table~\ref{tab:tails} are consistent with $3/d_O$
for the three shortest paths, and the value for the DFS branch is about two
standard errors above it.  The left tails thus decay algebraically, in
contrast to the decay $\exp(-bx^{-2/d_f})$ of the largest cluster of critical
percolation and of the Fortuin--Kasteleyn representation of the Ising model,
where $d_f$ is the fractal dimension of the cluster
\cite{ZhuEtAl2026Tails}.  The right tails in panels (a2)--(d2) agree with
Eq.~\eqref{eq:path-right-tail}, and the data for different $L$ coincide.
Fits with free $\beta_O$ give the values in Table~\ref{tab:tails}, which agree
with $2/(2-d_O)$ within the errors.

\section{Discussion}
\label{sec:discussion}

In this work, we studied a family of paths on the trace of simple random
walks on the infinite cubic lattice, which join the starting site to the
final position of the walker.  First, different path rules lead to distinct
fractal dimensions.  Requiring the shortest path to follow the traversal
directions increases its dimension, and requiring in addition the time order
increases it further, so that the direction and the order of the traversals
change the scaling of the shortest path and not only its amplitude.  The
branch of a DFS tree, which is not chosen to be short, has the largest
dimension.  Second, the standard deviation of each path length grows with
the same exponent as its mean, so that the path lengths are not
self-averaging, in contrast to the mass, and their distributions collapse
onto scaling functions of finite width.  Finally, the left and right tails of
these functions are consistent with $F_{O,L}(x)\sim x^{3/d_O}$ and
$\exp(-cx^{2/(2-d_O)})$, respectively.

These findings show that the way in which a path is selected on the trace affects its fractal dimension.  The allowed paths of $S_1$, $S_2$, and
$S_3$ form nested sets, and each of them takes the minimum length over its
set.  The LERW is also a time-respecting path, but it is fixed by the history
of the walk and not minimized, so that $d_{\rm LE}>d_{S_3}$
\cite{Wilson2010LERW}.  The DFS branch is self-avoiding, and its dimension is
slightly larger than that of the self-avoiding walk, although it is selected
by the search rule on the trace and not with the uniform weight of the
self-avoiding walk.  No analytic theory is available for these exponents.
An analytic calculation would show how the fractal dimension of a path
depends on the path rule, and rigorous bounds could establish their
ordering.

The study of these paths can be extended to other dimensions, and two
dimensions are of particular interest because of the recurrence threshold at
$d=2$.  Here, the expected number of returns grows logarithmically with the
number of steps $N$, and the mass scales as $N/\ln N$
\cite{LawlerLimic2010,DvoretzkyErdos1951}.  The question is whether the
shortest-path and DFS lengths retain pure power laws or acquire logarithmic
factors.  Connecting these scaling laws to the returns and the path
intersections could help explain how the path geometry depends on the
spatial dimension.

\section*{Acknowledgment}
Q.S. and Y.D. were supported by the National Natural Science Foundation of
China under Grant No.~12275263 and the Quantum Science and Technology
National Science and Technology Major Project under Grant No.~2021ZD0301900.
X.L. was supported by the National Key R\&D Program of China under Grant
No.~2021YFA1002700 and the Beijing Natural Science Foundation under Grant
No.~JQ26001. X.L. also thanks Ewain Gwynne for proposing to investigate
the time-respecting shortest path exponent $S_3$.

\bibliography{references}

\end{document}